\documentstyle[12pt,epsf]{article}
\newcommand{\n}{\hspace*{-2.5mm}}
\newcommand{\gsim}{\;\rlap{\lower 3.5 pt \hbox{$\mathchar \sim$}} \raise 1pt
 \hbox {$>$}\;}
\newcommand{\lsim}{\;\rlap{\lower 3.5 pt \hbox{$\mathchar \sim$}} \raise 1pt
 \hbox {$<$}\;}

\begin{document}


\title{\vskip-3cm{\baselineskip14pt
\centerline{\normalsize\hfill MPI/PhT/96--110}
\centerline{\normalsize\hfill October 1996}
}
\vskip1.5cm
Three-Loop ${\cal O}(\alpha_s^2G_FM_t^2)$ Corrections to Hadronic Higgs
Decays}
\author{{\sc K.G. Chetyrkin}\thanks{Permanent address:
Institute for Nuclear Research, Russian Academy of Sciences,
60th October Anniversary Prospect 7a, Moscow 117312, Russia.},
{\sc B.A. Kniehl, and M. Steinhauser}\\
{\normalsize Max-Planck-Institut f\"ur Physik (Werner-Heisenberg-Institut),}\\
{\normalsize F\"ohringer Ring 6, 80805 Munich, Germany}\\ \\
}
\date{}
\maketitle
\begin{abstract}
We calculate the top-quark-induced three-loop corrections of
${\cal O}(\alpha_s^2G_FM_t^2)$ to the Yukawa couplings of the first five quark 
flavours in the framework of the minimal standard model with an 
intermediate-mass Higgs boson, with mass $M_H\ll2M_t$.
The calculation is performed using an effective-Lagrangian approach
implemented with the hard-mass procedure.
As an application, we derive the ${\cal O}(\alpha_s^2G_FM_t^2)$ corrections to
the $H\to q\bar q$ partial decay widths, including the case $q=b$.
The couplings of the Higgs boson to pairs of leptons and intermediate
bosons being known to ${\cal O}(\alpha_s^2G_FM_t^2)$, this completes the
knowledge of such corrections in the Higgs sector.
We express the results both in the $\overline{\mbox{MS}}$ and on-shell schemes
of mass renormalization.
We recover the notion that the QCD perturbation expansions exhibit a worse
convergence behaviour in the on-shell scheme than they do in the
$\overline{\mbox{MS}}$ scheme.

\end{abstract}
\newpage


\section{Introduction}

One of the longstanding questions of elementary particle physics is whether
nature makes use of the Higgs mechanism of spontaneous symmetry breaking to
endow the particles with their masses.
The Higgs boson, $H$, is the missing link sought to verify this theoretical
conjecture in the standard model (SM).
The possible range of the Higgs-boson mass, $M_H$, is constrained from below
both experimentally and theoretically.
The failure of experiments at the CERN Large Electron-Positron Collider
(LEP~1) to observe the decay $Z\rightarrow f\bar f H$ has ruled out the mass
range $M_H\le65.6$~GeV at the 95\% confidence level \cite{jan}.
Depending on the precise value of the top-quark mass, $M_t$, the requirement
that the vacuum be the true ground state provides an even more stringent
theoretical lower bound \cite{lin}.
However, this bound may be somewhat relaxed by taking into account the
possibility that the physical minimum of the effective potential might be
metastable \cite{esp}.
Other theoretical arguments bound $M_H$ from above.
The requirements that partial-wave unitarity in intermediate-boson scattering
at high energies be satisfied \cite{dic} or that perturbation theory in the
Higgs sector be meaningful \cite{vel} establish an upper bound on $M_H$ at
about $\left(8\pi\sqrt2/3G_F\right)^{1/2}\approx1$~TeV, where $G_F$ denotes
Fermi's constant, in a weakly interacting SM.
The triviality bounds, {\it i.e.}, $M_H$ upper bounds derived through
perturbative \cite{cab} or lattice \cite{has} computations by requiring that
the running Higgs self-coupling, $\lambda(\mu)$, stay finite for
renormalization scales $\mu<\Lambda$, where $\Lambda$ is the cutoff beyond 
which new physics operates, are somewhat stronger.
In the following, we shall focus our attention on the lower end of the allowed
$M_H$ range, where $M_H\ll2M_t$, which will be accessed by colliding-beam 
experiments in the near future. 

A Higgs boson with $M_H\lsim135$~GeV decays dominantly to $b\bar b$ pairs
\cite{pre}.
This decay mode will be of prime importance for Higgs-boson searches at
LEP~2 \cite{gkw}, the Fermilab Tevatron \cite{sta} or a possible 4-TeV upgrade
thereof \cite{mar}, a next-generation $e^+e^-$ linear collider \cite{imh},
and a future $\mu^+\mu^-$ collider \cite{bar}.
Techniques for the measurement of the $H\to b\bar b$ branching fraction at a
$\sqrt s=500$~GeV $e^+e^-$ linear collider have been elaborated in
Ref.~\cite{hil}.
The branching ratio of $H\to c\bar c$ is roughly 30 times smaller than that of
$H\to b\bar b$ \cite{pre}.

Once a novel scalar particle is discovered, it will be crucial to decide if it
is the very Higgs boson of the SM or if it lives in some more extended Higgs
sector.
To that end, precise knowledge of the SM predictions will be mandatory,
{\it i.e.}, quantum corrections must be taken into account.
The present knowledge of quantum corrections to the $H\to q\bar q$ partial
decay widths has recently been reviewed in Ref.~\cite{mod}.
At one loop, the electroweak \cite{fle,bak} and quantum-chromodynamical (QCD)
\cite{bra} corrections are known for arbitrary masses.
The leading high-$M_H$ term, of ${\cal O}(G_FM_H^2)$, was first derived by
Veltman \cite{pol}.
In the limit $M_H\ll2M_t$, which we are interested in here, the terms of
${\cal O}(X_t)$, where $X_t=\left(G_FM_t^2/8\pi^2\sqrt2\right)$, tend to be
dominant.
They arise in part from the renormalizations of the Higgs wave function and
vacuum expectation value, which are independent of the quark flavour $q$ 
\cite{cha}.
In the case of $q=b$, there is an additional non-universal ${\cal O}(X_t)$
contribution \cite{fle,bak,yao}, which partly cancels the flavour-independent
one.
At two loops, the universal \cite{hll} and bottom-specific \cite{hbb,kwi}
${\cal O}(\alpha_sX_t)$ terms are available.
Furthermore, the first \cite{gor} and second \cite{sur} terms of the
expansion in $M_q^2/M_H^2$ of the ${\cal O}(\alpha_s^2)$ five-flavour QCD
correction have been found.
As for the top-quark-induced correction in ${\cal O}(\alpha_s^2)$, the full
$M_t$ dependence of the non-singlet (double-bubble) contribution \cite{kni} as
well as the first four terms of the $M_H^2/M_t^2$ expansion of the singlet
(double-triangle) contribution \cite{lar} have been computed.
At three loops, the ${\cal O}(\alpha_s^3)$ non-singlet correction is known in
the massless approximation \cite{che}.

In this paper, we shall take the next step, to three loops including virtual
top-quark effects.
Specifically, we shall evaluate the ${\cal O}(\alpha_s^2X_t)$ corrections to
the decay widths of $H\to q\bar q$, and in particular of $H\to b\bar b$.
These corrections may be divided into three classes, which are separately 
finite and gauge independent:
$(i)$ The {\it universal} correction originates in the renormalizations of the
Higgs wave function and vacuum expectation value and occurs as a building
block in the computation of quantum corrections to any Higgs-boson production
or decay process.
It is related to the Higgs- and $W$-boson self-energies.
In the case of the leptonic decay, $H\to l^+l^-$, this is the only source of
${\cal O}(\alpha_s^2X_t)$ corrections \cite{ks}.
$(ii)$ The {\it quark-specific} correction arises from Feynman diagrams where
the top quark only appears in a closed loop which is connected with the
external quark line by two gluons (and, in some cases, one additional weak
neutral boson).
It just depends on the third component of weak isospin of the considered 
quark.
$(iii)$ The {\it bottom-specific} correction emerges from the one-loop seed 
diagram where one charged boson is emitted and re-absorbed from the external
bottom-quark line, by appropriately adding gluon lines (and, in some cases,
one quark or ghost loop).
It is the natural extension of the bottom-specific ${\cal O}(\alpha_sX_t)$
correction to the $H\to b\bar b$ decay width \cite{hbb,kwi} by one order of
$\alpha_s$.

The next-to-leading-order QCD corrections to the top-quark-induced shifts in 
the $l^+l^-H$, $W^+W^-H$, and $ZZH$ couplings have been found in
Ref.~\cite{ks}.
The present paper completes the knowledge of the ${\cal O}(\alpha_s^2X_t)$ 
corrections to the SM Yukawa couplings, still excluding that of the top quark.
The Yukawa couplings of the first four quark flavours receive corrections from
the same class of diagrams, while, in the case of the bottom Yukawa coupling,
an additional class of diagrams must be included.

In Ref.~\cite{ks}, it was noticed that the QCD perturbation expansions of the
$l^+l^-H$, $W^+W^-H$, and $ZZH$ couplings and the electroweak $\rho$ 
parameter, for which the ${\cal O}(\alpha_s^2X_t)$ correction is also known
\cite{avd}, exhibit striking similarities.
If the top-quark mass is renormalized according to the on-shell scheme, the
coefficients of $(\alpha_s/\pi)X_t$ and $(\alpha_s/\pi)^2X_t$ are in each
case negative and of increasing magnitude.
On the other hand, the corresponding coefficients in the
$\overline{\mbox{MS}}$ scheme are much smaller and of variant sign.
This gave support to the notion that the use of the pole mass deteriorates the
convergence properties of the QCD perturbation series, which may also be
motivated from the study of renormalons \cite{ren}.
It is interesting to find out whether this observation is substantiated by the
analysis of the light-quark and bottom Yukawa couplings.
We shall return to this issue in Sect.~\ref{sechbb}.

Our key result for the $H\to b\bar b$ decay width to ${\cal O}(\alpha_s^2X_t)$ 
has recently been presented without derivation in a brief note \cite{cks}.
This paper provides the full details of our analysis and also deals with the
$H\to q\bar q$ decays, where $q=u,d,s,c$.
It is organized as follows.
After introducing our notations in Sect.~\ref{secnot}, we shall derive, in
Sect.~\ref{seceff}, a heavy-top-quark effective Lagrangian with two
coefficient functions to be determined by diagrammatic calculation.
From this Lagrangian, we shall obtain a generic formula for the $H\to q\bar q$
decay widths valid through ${\cal O}(\alpha_s^2X_t)$.
In Sect.~\ref{sechqq}, we shall calculate, at three loops, the coefficient
functions relevant for the first four quark flavours.
In Sect.~\ref{sechbb}, we shall extend this analysis to also include the case
$q=b$.
Our conclusions will be summarized in Sect.~\ref{seccon}.


\section{Notations}
\label{secnot}

In this section, the notation is fixed and useful formulae, which will be
necessary to numerically evaluate our results, are provided.
The calculation is performed in the framework of dimensional 
regularization with space-time dimension $n=4-2\varepsilon$.
The QCD gauge group is taken to be SU($N_c$), with $N_c$ arbitrary.
The colour factors corresponding to the Casimir operators of the fundamental
and adjoint representations are $C_F=(N_c^2-1)/(2N_c)$ and $C_A=N_c$, 
respectively.
For the numerical evaluation we set $N_c=3$.
The trace normalization of the fundamental representation is $T=1/2$.
The number of active quark flavours is denoted by $n_f$.
Unless otherwise stated, we work in the $\overline{\mbox{MS}}$ scheme, with
$\mu$ being the renormalization scale.

The $\mu$ dependence of the strong coupling constant, $\alpha_s(\mu)$, and
any quark mass, $m(\mu)$, is governed by the renormalization-group (RG)
equations,
\begin{eqnarray}
\label{rgealpha}
\frac{\mu^2d}{d\mu^2}\,\frac{\alpha_s(\mu)}{\pi}
&\n=\n& \beta(\alpha_s)= 
-\left(\frac{\alpha_s}{\pi}\right)^2\sum_{k\ge0}\beta_k
                 \left(\frac{\alpha_s}{\pi}\right)^k,
\\
\frac{\mu^2d}{d\mu^2}\ln m(\mu)
&\n=\n&\gamma_{m}(\alpha_s)=
-\frac{\alpha_s}{\pi}
\sum_{k\ge0}\gamma_{m}^k  \left(\frac{\alpha_s}{\pi}\right)^k,
\end{eqnarray}
where
\begin{eqnarray}
\label{beta}
\beta_0 
  =  
\frac{1}{4}\left[
\frac{11}{3} C_A - \frac{4}{3} T n_f \right],
&&
\beta_1 
 =  
\frac{1}{16}\left[
\frac{34}{3} C_A^2 - 4 C_F T n_f
- \frac{20}{3} C_A T n_f \right],
\\
\gamma_{m}^0 
 =  
\frac{1}{4}\bigg[3 C_F\bigg],
&&
\gamma_{m}^1 
 =  
\frac{1}{16}\left[
\frac{3}{2} C_F^2 + \frac{97}{6} C_F C_A 
- \frac{10}{3} C_F T n_f \right]
\end{eqnarray}
are the first few coefficients of the QCD $\beta$ and $\gamma$ functions.
The next-to-leading-order (two-loop) solution of Eq.~(\ref{rgealpha}) reads
\begin{eqnarray}
\frac{\alpha_s(\mu)}{\pi} &\n=\n&
\frac{1}{\beta_0\ln(\mu^2/\Lambda^2)}\left[1-\frac{\beta_1}{\beta_0^2}\,
                          \frac{\ln\ln(\mu^2/\Lambda^2)}{\ln(\mu^2/\Lambda^2)}
                  \right],
\end{eqnarray}
where $\Lambda$ is the asymptotic scale parameter.

The relation between the $\overline{\mbox{MS}}$ mass $m(\mu)$ and the on-shell
mass $M_t$ of the top quark is given by
\begin{eqnarray}
\frac{m_t(\mu)}{M_t} &\n=\n&
1
+ X_t\left(4+\frac{3}{2}\ln\frac{\mu^2}{M_t^2}\right)
+ \frac{\alpha_s^{(n_f)}(\mu)}{\pi}C_F\left(-1
              -\frac{3}{4}\ln\frac{\mu^2}{M_t^2}\right)
\nonumber\\
&\n\n&
{}+ \left(\frac{\alpha_s^{(n_f)}(\mu)}{\pi}\right)^2 \left\{
  C_F T \left(
            \frac{3}{4}
          - \frac{3}{2}\zeta(2)
        \right)\right.
\nonumber\\
&\n\n&
{}+C_F^2 \left[
            \frac{7}{128}
          + 3\zeta(2)\left(-\frac{5}{8}+\ln2\right)
          - \frac{3}{4}\zeta(3)
          + \frac{21}{32}\ln\frac{\mu^2}{M_t^2}
          + \frac{9}{32}\ln^2\frac{\mu^2}{M_t^2}
        \right]
\nonumber\\
&\n\n&
{}+ C_A C_F \left[
          - \frac{1111}{384}
          + \frac{1}{2}\zeta(2)(1-3\ln2)
          + \frac{3}{8}\zeta(3)
          - \frac{185}{96}\ln\frac{\mu^2}{M_t^2}
          - \frac{11}{32}\ln^2\frac{\mu^2}{M_t^2}
          \right]
\nonumber\\
&\n\n&{}+\left. C_F T n_f \left(
            \frac{71}{96}
          + \frac{1}{2}\zeta(2)
          + \frac{13}{24}\ln\frac{\mu^2}{M_t^2}
          + \frac{1}{8}\ln^2\frac{\mu^2}{M_t^2}
            \right)
\right\}+{\cal O}(\alpha_s^3,\alpha_sX_t,X_t^2),
\label{eqmbarmos}
\end{eqnarray}
where $n_f=6$, $X_t=\left(G_F M_t^2/8\pi^2\sqrt2\right)$, and $\zeta$ is
Riemann's zeta function, with values $\zeta(2)=\pi^2/6$ and
$\zeta(3)\approx1.202\,057$.
The ${\cal O}(\alpha_s^2)$ and ${\cal O}(X_t)$ corrections were
calculated in Refs.~\cite{GraBroGraSch90} and \cite{hem,boc}, respectively.
Iterating Eq.~(\ref{eqmbarmos}), one obtains the scale-independent
$\overline{\mbox{MS}}$ mass $\mu_t=m_t(\mu_t)$ as
\begin{eqnarray}
\frac{\mu_t}{M_t} &\n=\n& 
1
+ 4 X_t
-\frac{\alpha_s^{(n_f)}(M_t)}{\pi}C_F
+\left(\frac{\alpha_s^{(n_f)}(M_t)}{\pi}\right)^2\left\{
C_FT\left(\frac{3}{4}-\frac{3}{2}\zeta(2)\right)\right.
\nonumber\\
&\n\n&{}+
C_F^2\left[\frac{199}{128}+3\zeta(2)\left(-\frac{5}{8}+\ln2\right)
-\frac{3}{4}\zeta(3)\right]
+C_AC_F\left[
-\frac{1111}{384}
\right.
\nonumber\\
&\n\n&{}+
\left.\left.
\frac{1}{2}\zeta(2)(1-3\ln2)+\frac{3}{8}\zeta(3)\right]
+C_FTn_f\left(\frac{71}{96}+\frac{1}{2}\zeta(2)\right)
\right\}+{\cal O}(\alpha_s^3,\alpha_sX_t,X_t^2).
\end{eqnarray}
In the $\overline{\mbox{MS}}$ scheme, the relation between the values of
$\alpha_s(\mu)$ for $n_f=5$ and $n_f=6$ is given by 
\begin{equation}
\label{eqmatch}
\alpha_s^{(5)}(\mu)
= \alpha_s^{(6)}(\mu)\left\{
      1 + \frac{\alpha_s^{(6)}(\mu)}{\pi}T
        \left[-\frac{1}{3}\ln\frac{\mu^2}{m_t^2}
      + x_t
        \left(-\frac{2}{3} + \ln\frac{\mu^2}{m_t^2}\right)
        \right]+{\cal O}(\alpha_s^2)\right\},
\end{equation}
where $x_t(\mu)=\left[G_F m_t^2(\mu)/8\pi^2\sqrt{2}\right]$.
The constant term of ${\cal O}(\alpha_s^2x_t)$ in Eq.~(\ref{eqmatch})
represents a new result, which follows from the analysis of the next section.
Inserting the ${\cal O}(X_t)$ term of Eq.~(\ref{eqmbarmos}) into
Eq.~(\ref{eqmatch}), one obtains
\begin{equation}
\label{eqmatchos}
\alpha_s^{(5)}(\mu)
=\alpha_s^{(6)}(\mu)\left\{
      1 + \frac{\alpha_s^{(6)}(\mu)}{\pi}T
        \left[-\frac{1}{3}\ln\frac{\mu^2}{M_t^2}
      + X_t
        \left( 2+2\ln\frac{\mu^2}{M_t^2} \right)
        \right]+{\cal O}(\alpha_s^2)\right\}.
\end{equation}


\section{Effective Lagrangian}
\label{seceff}

In this section, we construct an effective Lagrangian for the interaction
between an intermediate-mass Higgs boson and a quark-antiquark pair.
Therefore, in addition to the pure QCD Lagrangian, the couplings of the quarks
to the Higgs boson, $H$, the neutral Goldstone boson, $\chi$, and
the charged Goldstone bosons, $\phi^\pm$, must be taken into account.
This will produce corrections proportional to $X_t$.
In this paper, we are not interested in corrections of ${\cal O}(X_t^2)$, and
our formulae will not in general be valid in this order.

As a starting point, we consider the bare Yukawa Lagrangian,
\begin{equation}
{\cal L}_Y = -\frac{H^0}{v^0}J,
\end{equation}
where $v$ is the Higgs vacuum-expectation value,
the superscript 0 labels bare quantities,
and the operator $J$ is defined as
\begin{equation}
J = \sum_q m_q^0 \bar{q}^0 q^0 + m_t^0 \bar{t}^0 t^0.
\end{equation}
Here, $q$ runs over $u$, $d$, $s$, $c$, and $b$.
It is easy to see that
$J$ is a finite operator, in the sense that no additional renormalization
constant is needed.
Our aim is to construct the equivalent expression for $J$
in the effective theory where the top quark is integrated out.
Because only the leading terms in $M_t$ are considered, ${\cal L}_Y$ may
be written as a linear combination of three physical operators
with mass dimension four, namely,
\begin{equation}
\label{leff}
{\cal L}_Y \stackrel{m_t^0\to\infty}{\longrightarrow}
{\cal L}_Y^{\rm eff} = -\frac{H^0}{v^0}\left[
C_1^0O_1^\prime+\sum_q\left(
C_{2q}^0O_{2q}^\prime+C_{3q}^0O_{3q}^\prime\right)\right],
\end{equation}
where
\begin{eqnarray}
O_1^\prime&\n=\n&\left(G_{a\mu\nu}^{0\prime}\right)^2,\nonumber\\
O_{2q}^\prime&\n=\n&m_q^{0\prime}\bar q^{0\prime}q^{0\prime},\nonumber\\
O_{3q}^\prime&\n=\n&\bar q^{0\prime}\left[\frac{i}{2}
\left(
\stackrel{\rightarrow}{D}\hspace{-.85em}/{}^{0\prime}
-\stackrel{\leftarrow}{D}\hspace{-.85em}/{}^{0\prime}
\right)
-m_q^{0\prime}\right]q^{0\prime}
\end{eqnarray}
are bare operators in the $n_f=5$ effective theory and
$C_i^0 = C_i^0(\alpha_s^0,m_t^0,\mu)$ ($i=1,2q,3q$) are their bare coefficient
functions.
Here, $G_{a\mu\nu}$ is the colour field strength, $D_\mu$ is the covariant
derivative, and the parameters and fields of the effective theory are
marked by a prime.
Notice that $O_{3q}^\prime$ vanishes by the fermionic equation of motion and
may be omitted once $C_1^0$ and $C_{2q}^0$ are determined.

The relations between the parameters and fields in the full and effective
theories read
\begin{eqnarray}
q^{0\prime} &\n =\n & \left(\zeta_{2q}^0\right)^{1/2} q^0,\nonumber\\
G_{a\mu}^{0\prime} &\n=\n& \left(\zeta_3^0\right)^{1/2} G_{a\mu}^0,\nonumber\\
m_q^{0\prime} &\n=\n& \zeta_{m,q}^0 m_q^0,\nonumber\\
g_s^{0\prime} &\n=\n& \zeta_g^0 g_s^0,
\end{eqnarray}
where $G_{a\mu}$ denotes the gluon field.
The renormalization constants $\zeta_{2q}^0$, $\zeta_{m,q}^0$, and $\zeta_3^0$
play an important r\^ole in the determination of the coefficient functions
$C_i^0$.
As may be seen using the method of projectors \cite{GorLar87}, they may be
calculated from the vector and scalar parts of the quark self-energy,
$\Sigma_V(p^2)$ and $\Sigma_S(p^2)$, and the transverse part of the gluon
self-energy, $\Pi(p^2)$, via
\begin{eqnarray}
\zeta_{2q}^0 &\n=\n& 1+\Sigma_V^{0t}(0),
\label{eqzeta2}
\\
\zeta_{m,q}^0 &\n=\n& \frac{1-\Sigma_S^{0t}(0)}{1+\Sigma_V^{0t}(0)},
\label{eqzetam}
\\
\zeta_3^0 &\n=\n&1+\Pi^{0t}(0),
\label{eqzeta3}
\end{eqnarray}
where the superscript $t$ indicates that only diagrams containing the top
quark have to be considered and the superscript $0$ reminds us of that the
parameters are still in their bare forms.
In our convention, the bare quark and gluon propagators are proportional to 
$\left[\not\!p\left(1+\Sigma_V^0(p^2)+\gamma_5\Sigma_A^0(p^2)\right)
-m_q^0\left(1-\Sigma_S^0(p^2)\right)\right]^{-1}$
and $[p^2(1+\Pi^0(p^2))]^{-1}$, respectively.
Notice that the axial-vector part of the quark self-energy, $\Sigma_A(p^2)$,
does not enter our analysis because it leads to corrections of
${\cal O}(x_t^2)$.
The renormalization constant $\zeta_g^0$ only enters the stage via
Eqs.~(\ref{eqmatch}) and (\ref{eqmatchos}).
Through ${\cal O}(\alpha_s^2x_t)$, we have
$\zeta_g^0=\left(\zeta_3^0\right)^{-1/2}$.

By means of the higher-order formulation \cite{spi} of a well-known
low-energy theorem (LET) \cite{let} in connection with the method of
projectors \cite{GorLar87}, one may derive the following relations which allow
one to calculate the coefficient functions:
\begin{eqnarray}
\zeta_3^0C_1^0&\n=\n&-\frac{m_t^0\partial}{\partial m_t^0}\Pi^{0t}(0),
\label{eqc01}
\\
\zeta_{2q}^0\zeta_{m,q}^0\left(C_{2q}^0-C_{3q}^0\right)&\n=\n&
1-\Sigma_S^{0t}(0)-\frac{m_t^0\partial}{\partial m_t^0}\Sigma_S^{0t}(0),
\label{eqc2c3}
\\
\zeta_{2q}^0C_{3q}^0&\n=\n& 
-\frac{m_t^0\partial}{\partial m_t^0}\Sigma_V^{0t}(0).
\label{eqc3}
\end{eqnarray}
Here it is understood that the operator $(m_t^0\partial/\partial m_t^0)$ only
acts on those appearances of $m_t^0$ which remain after every $1/v^0$ is
saturated by one power of $m_t^0$ to give $m_t^0/v^0$.
As may be seen from Eqs.~(\ref{eqc01})--(\ref{eqc3}),
the calculation of the coefficient functions involves
the quark- and gluon-propagator diagrams containing 
the top quark, with nullified external momenta.
By means of the LET, the corresponding vertex diagrams may be generated by
attaching an external Higgs-boson line with zero momentum to the quark
lines.
To the order of interest here, this requires the evaluation of one-, two-, and
three-loop tadpole diagrams.
They are calculated with the help of the program package MATAD, written in
FORM \cite{FORM}, which makes use of integration-by-parts identities developed
in Ref.~\cite{Bro92}.

So far, we have constructed an expression for $J$ in terms of bare operators
and coefficient functions, which both still contain poles in
$\varepsilon$.
As is well known, different operators of the same dimension and quantum 
numbers in general mix under renormalization.
Specifically, the renormalized operators, which will be denoted by square
brackets, are related to the unrenormalized ones according to \cite{anomdim}
\begin{eqnarray}
\left[O_1^\prime\right]&\n=\n&
\left[1+2\left(\frac{\alpha_s\partial}{\partial\alpha_s}\ln Z_g\right)\right]
O_1^\prime
-4\left(\frac{\alpha_s\partial}{\partial\alpha_s}\ln Z_m\right)
\sum_q O_{2q}^\prime,
\nonumber\\
\left[O_{2q}^\prime\right]&\n=\n&O_{2q}^\prime,
\end{eqnarray}
where $Z_g$ and $Z_m$ are the coupling and mass renormalization constants,
respectively, in pure QCD with $n_f=5$ active flavours.
It hence follows that
\begin{eqnarray}
\frac{\mu^2d}{d\mu^2}\left[O_1^\prime\right]&\n=\n&
-\left(\frac{\alpha_s\partial}{\partial\alpha_s}\,
\frac{\pi\beta}{\alpha_s}\right)\left[O_1^\prime\right]
+4\left(\frac{\alpha_s\partial}{\partial\alpha_s}\gamma_m\right)
\sum_q\left[O_{2q}^\prime\right],
\nonumber\\
\frac{\mu^2d}{d\mu^2}\left[O_{2q}^\prime\right]&\n=\n&0.
\end{eqnarray}
Note that $O_{3q}^\prime$ does not mix with $O_1^\prime$ and $O_{2q}^\prime$.
On the other hand, the coefficient functions are renormalized according to
\begin{eqnarray}
C_1&\n=\n&\frac{1}{1+2(\alpha_s\partial/\partial\alpha_s)\ln Z_g}C_1^0,
\nonumber\\
C_{2q}&\n=\n&\frac{4(\alpha_s\partial/\partial\alpha_s)\ln Z_m}
{1+2(\alpha_s\partial/\partial\alpha_s)\ln Z_g}C_1^0+C_{2q}^0.
\end{eqnarray}     
Consequently, the $n_f=5$ effective Lagrangian~(\ref{leff}) takes the form
\begin{equation}
{\cal L}_Y^{\rm eff} = -\frac{H^0}{v^0}
\left[C_1\left[O_1^\prime\right]
+\sum_q\left(C_{2q}\left[O_{2q}^\prime\right]
+C_{3q}\left[O_{3q}^\prime\right]\right)
\right].
\label{eqleff}
\end{equation}
The new coefficient functions and the operators are individually finite,
but, with the exception of $\left[O_{2q}^\prime\right]$, they are not
separately RG invariant.
From now on, $\left[O_{3q}^\prime\right]$ will be omitted because it vanishes
on mass shell and thus does not contribute to the $H\to q\bar q$ decay widths.

It is desirable to arrange for the renormalized coefficient functions 
and operators to be separately $\mu$ independent up to higher orders.
This may be achieved by reshuffling the terms in Eq.~(\ref{eqleff}).
Exploiting the RG invariance of $\left[O_{2q}^\prime\right]$ and of the trace
of the energy-momentum tensor,
\begin{equation}
\left[\Theta_{\mu}^{\mu}\right] 
= \frac{\pi\beta^{(5)}}{2\alpha_s^{(5)}} \left[O_1^\prime\right] +
\left(1 - 2\gamma_m^{(5)}\right)\sum_q\left[O_{2q}^\prime\right],
\end{equation}
where all quantities are defined in the $n_f=5$ effective theory, we may
construct two new operators which are indeed RG invariant
\cite{anomdim2}, {\it e.g.},
\begin{eqnarray}
\left[O_g^\prime\right] &\n=\n& - \frac{2\pi}{\beta_0^{(5)}}\left(
         \frac{\pi\beta^{(5)}}{2\alpha_s^{(5)}}\left[O_1^\prime\right]
        -2\gamma_m^{(5)}\sum_q\left[O_{2q}^\prime\right]
                                \right),
\nonumber\\
{\left[O_q^\prime\right]} &\n=\n& \left[O_{2q}^\prime\right].
\end{eqnarray}
Thus, the relevant part of the $n_f=5$ effective Lagrangian may be written as
\begin{equation}
{\cal L}_Y^{\rm eff} = 
-\frac{H^0}{v^0}\left(C_g\left[O_g^\prime\right]
+ \sum_qC_q\left[O_q^\prime\right]\right),
\label{eqleff2}
\end{equation}
where
\begin{eqnarray}
C_g &\n=\n& -\frac{\alpha_s^{(5)}\beta_0^{(5)}}{\pi^2\beta^{(5)}}C_1,
\nonumber\\
C_q &\n=\n& \frac{4\alpha_s^{(5)}\gamma_m^{(5)}}{\pi\beta^{(5)}}C_1+C_{2q}.
\end{eqnarray}
As the new coefficient functions and operators are separately RG invariant,
we may choose $\mu=\mu_t$ for $C_g$ and $C_q$ and $\mu=M_H$ for
$\left[O_g^\prime\right]$ and $\left[O_q^\prime\right]$.

In order to calculate the $H\to q\bar q$ decay width, it is more convenient
to re-express ${\cal L}_Y^{\rm eff}$ in terms of the
operators $\left[O_1^\prime\right]$ and $\left[O_{2q}^\prime\right]$.
At the same time, the separation of the scales $M_H$ and $\mu_t$ is kept,
so that Eq.~(\ref{eqleff2}) becomes
\begin{equation}
{\cal L}_Y^{\rm eff} = 
-\frac{H^0}{v^0}\left({\cal C}_1 \left[O_1^\prime\right] +
\sum_q{\cal C}_{2q} \left[O_{2q}^\prime\right] \right),
\label{eqlefffin}
\end{equation}
where
\begin{eqnarray}
{\cal C}_1(\mu_t,M_H) &\n=\n& \frac{\alpha_s^{(5)}(\mu_t)\beta^{(5)}(M_H)}
{\alpha_s^{(5)}(M_H)\beta^{(5)}(\mu_t)}C_1(\mu_t),
\nonumber\\
{\cal C}_{2q}(\mu_t,M_H) &\n=\n&
\frac{4\alpha_s^{(5)}(\mu_t)}{\pi\beta^{(5)}(\mu_t)}
\left(\gamma_m^{(5)}(\mu_t)-\gamma_m^{(5)}(M_H)\right)C_1(\mu_t)
+C_{2q}(\mu_t).
\label{eqctil}
\end{eqnarray}
Notice that, for our purposes, ${\cal C}_1$ and ${\cal C}_{2q}$ are needed up
to ${\cal O}(\alpha_s x_t)$ and ${\cal O}(\alpha_s^2 x_t)$, respectively.
It is instructive to expand Eq.~(\ref{eqctil}) in terms of
$\alpha_s^{(6)}(\mu_t)$, keeping only terms relevant for our
${\cal O}(\alpha_s^2 x_t)$ calculation.
Observing that $C_1$ starts at ${\cal O}(\alpha_s)$, while $C_{2q}=1$ to
lowest order, this yields
\begin{eqnarray}
\label{calc}
{\cal C}_1(\mu_t,M_H) &\n=\n& C_1(\mu_t),
\nonumber\\
{\cal C}_{2q}(\mu_t,M_H) &\n=\n& -4\gamma^0_m\frac{\alpha_s^{(6)}(\mu_t)}{\pi}
                            C_1(\mu_t)\ln\frac{\mu_t^2}{M_H^2} + C_{2q}(\mu_t).
\end{eqnarray}

The coefficients $C_1$ and $C_{2q}$ must be computed diagrammatically.
This will be done for the light-quark flavours in Sect.~\ref{sechqq} and for
$q=b$ in Sect.~\ref{sechbb}.
In the remainder of this section, we present a generic formula, derived from
Eq.~(\ref{eqlefffin}), for the $H\to q\bar q$ decay width, appropriate for
$M_H\ll2M_t$, which accommodates all presently known corrections.
It reads
\begin{eqnarray}
\label{gammahqq}
\Gamma\left(H\to q\bar q\right)&\n=\n&\Gamma_{q\bar q}^{\rm Born}
\left(1+\bar\delta_{\rm u}\right)^2\left[\left(1+\Delta_q^{\rm QED}\right)
\left(1+\left.\Delta_q^{\rm weak}\right|_{x_t=0}\right)
\left(1+\Delta_q^{\rm QCD}\right)\left({\cal C}_{2q}\right)^2\right.
\nonumber\\
&\n\n&{}+\left.\Xi_q^{\rm QCD}{\cal C}_1{\cal C}_{2q}\right].
\end{eqnarray}
Here,
\begin{equation}
\label{born}
\Gamma_{q\bar q}^{\rm Born}=\frac{N_cG_FM_Hm_q^2}{4\pi\sqrt2}
\left(1-\frac{4m_q^2}{M_H^2}\right)^{3/2}
\end{equation}
is the Born result including the full mass dependence.
As is well known \cite{bra},
we may avoid the appearance of large logarithms of the type $\ln(M_H^2/m_q^2)$
in the QCD correction, $\Delta_q^{\rm QCD}$, by taking $m_q$ in
Eq.~(\ref{born}) to be the $\overline{\mbox{MS}}$ mass for $n_f=5$,
$m_q^{(5)}(\mu)$, with $\mu$ of order $M_H$.
Consequently, we may put $m_q=0$ in $\Delta_q^{\rm QCD}$.
We may proceed similarly with the quantum-electrodynamical (QED) correction,
$\Delta_q^{\rm QED}$, which then takes the form
\begin{equation}
\Delta_q^{\rm QED}=\frac{\alpha(\mu)}{\pi}Q_q^2
\left(\frac{17}{4}+\frac{3}{2}\ln\frac{\mu^2}{M_H^2}\right),
\end{equation}
where $\alpha(\mu)$ is the $\overline{\mbox{MS}}$ fine-structure constant
and $Q_q$ is the fractional quark charge.
In turn, $m_q^{(5)}(\mu)$ is then also shifted by a QED correction \cite{hem}
from the pole mass, $M_q$.
$\left.\Delta_q^{\rm weak}\right|_{x_t=0}$ denotes the weak correction with the
leading ${\cal O}(x_t)$ term stripped off.
If we put $m_q=0$ and consider the limit $M_H\ll2M_W$,
$\left.\Delta_q^{\rm weak}\right|_{x_t=0}$ simplifies to \cite{bak}
\begin{equation}
\left.\Delta_q^{\rm weak}\right|_{x_t=0}=\frac{G_FM_Z^2}{8\pi^2\sqrt2}
\left[\frac{1}{2}-3\left(1-4s_w^2|Q_q|\right)^2
+c_w^2\left(\frac{3}{s_w^2}\ln c_w^2-5\right)\right],
\end{equation}
where $c_w^2=1-s_w^2=M_W^2/M_Z^2$, $M_Z$ is the $Z$-boson mass, and
$M_W$ is the $W$-boson mass.
$\Delta_q^{\rm QCD}$ is the well-known QCD correction in the $n_f=5$ effective
theory \cite{gor},
\begin{eqnarray}
\Delta_q^{\rm QCD}&\n=\n&
\frac{\alpha_s^{(n_f)}(\mu)}{\pi} C_F
  \left(\frac{17}{4}+\frac{3}{2}\ln\frac{\mu^2}{M_H^2}\right)
\nonumber
\\
&\n\n&{}+
\left(\frac{\alpha_s^{(n_f)}(\mu)}{\pi}\right)^2
  \left[
         C_F^2\left( 
                \frac{691}{64} 
              - \frac{9}{4}\zeta(2)
              - \frac{9}{4}\zeta(3)
              + \frac{105}{16}\ln\frac{\mu^2}{M_H^2} 
              + \frac{9}{8}\ln^2\frac{\mu^2}{M_H^2} 
              \right)\right.
\nonumber\\
&\n\n&{}
  +      C_F C_A\left( 
                \frac{893}{64} 
              - \frac{11}{8}\zeta(2)
              - \frac{31}{8}\zeta(3)
              + \frac{71}{12}\ln\frac{\mu^2}{M_H^2} 
              + \frac{11}{16}\ln^2\frac{\mu^2}{M_H^2} 
              \right)
\nonumber\\
&\n\n&{}+\left.
         C_F T n_f \left( 
                - \frac{65}{16} 
                + \frac{1}{2}\zeta(2)
                + \zeta(3) 
                - \frac{11}{6}\ln\frac{\mu^2}{M_H^2} 
                - \frac{1}{4}\ln^2\frac{\mu^2}{M_H^2}
                   \right)\right]
\nonumber\\
&\n=\n&
\frac{\alpha_s^{(5)}(\mu)}{\pi}
  \left(\frac{17}{3}+2\ln\frac{\mu^2}{M_H^2}\right)
\nonumber
\\
&\n\n&{}+
\left(\frac{\alpha_s^{(5)}(\mu)}{\pi}\right)^2
  \left[
        \frac{8851}{144}
       -\frac{47}{6}\zeta(2)
       -\frac{97}{6}\zeta(3)
       +\frac{263}{9}\ln\frac{\mu^2}{M_H^2}
       +\frac{47}{12}\ln^2\frac{\mu^2}{M_H^2}
  \right].
\end{eqnarray}
This correction originates from the class of diagrams where the Higgs boson
directly couples to the final-state $q\bar q$ pair, {\it i.e.}, it does not
comprise the double-triangle topologies of Ref.~\cite{lar}.
The latter will be discussed below.
The scale $\mu$ in $\Delta_q^{\rm QCD}$ must be identified with that of
$m_q^{(5)}(\mu)$ in $\Gamma_{q\bar q}^{\rm Born}$.
It is natural to choose $\mu=M_H$ in order to suppress the logarithms of RG
origin.

The corrections of ${\cal O}(\alpha_s^nx_t)$, with $n=0,1,2$, as well as the
$n_f=6$ QCD corrections of ${\cal O}(\alpha_s)$ and ${\cal O}(\alpha_s^2)$ are
all contained in $\bar\delta_{\rm u}$, ${\cal C}_1$, and ${\cal C}_{2q}$.
$\bar\delta_{\rm u}$ contains the universal ${\cal O}(\alpha_s^nx_t)$
corrections, which originate from the renormalizations of the Higgs wave
function and vacuum expectation value.
Specifically, we have
\begin{equation}
\frac{H^0}{v^0}=2^{1/4}G_F^{1/2}H\left(1+\bar\delta_{\rm u}\right),
\end{equation}
with \cite{ks}
\begin{eqnarray}
\label{deltau}
\bar\delta_{\rm u} &\n=\n& x_t
\left\{ \frac{7}{2}
     + \frac{\alpha_s^{(6)}(\mu)}{\pi}
       \left(
             \frac{19}{3} 
           - 2\zeta(2)
           + 7\ln\frac{\mu^2}{m_t^2}
       \right)
     + \left(\frac{\alpha_s^{(6)}(\mu)}{\pi}\right)^2
       \left[
              \frac{13307}{864} 
            - \frac{2377}{108}\zeta(2)
\right.\right.
\nonumber\\
&\n\n&{}-
              \frac{178}{9}\zeta(3)
            + \frac{143}{12}\zeta(4)
            - \frac{1}{6} B_4 
            - \frac{1}{12} D_3
            + \frac{1323}{16} S_2
            + \left(\frac{267}{8}
            - \frac{15}{2}\zeta(2)\right)\ln\frac{\mu^2}{m_t^2}
\nonumber\\
&\n\n&{}+
\left.\left.
            \frac{105}{8}\ln^2\frac{\mu^2}{m_t^2}
        \right]
\right\},
\end{eqnarray}
where $S_2\approx0.260\,434$, $D_3\approx-3.027\,009$, and
$B_4\approx-1.762\,800$ are mathematical constants related to certain
three-loop tadpole diagrams.
The formula for $N_c$ arbitrary may be found in Ref.~\cite{ks}.
The choice $\mu=\mu_t$ eliminates the logarithms in Eq.~(\ref{deltau}) and may
thus be considered natural.

\begin{figure}[ht]
   \leavevmode
 \begin{center}
 \begin{tabular}{ccc}
\epsfxsize=5.0cm
\epsffile[189 314 423 478]{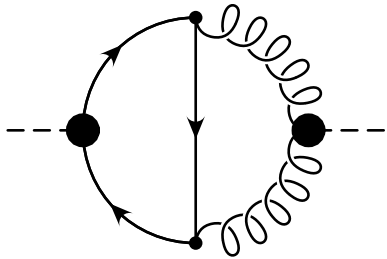}
 \end{tabular}
\caption{\label{figgammamix} Interference diagram of the operators
$\left[O_1^\prime\right]$ and $\left[O_{2q}^\prime\right]$, represented by the
solid circles, which generates $\Gamma_{q\bar q(g)}^{\rm mix}$.}
 \end{center}
\end{figure}

As is well known \cite{lar}, starting at ${\cal O}(\alpha_s^2)$,
$\Gamma\left(H\to q\bar q\right)$ also receives leading contributions from the
$q\bar q$ and $q\bar qg$ cuts of the double-triangle diagrams where the top
quark circulates in one of the triangles.
The additional exchange of a virtual Higgs or Goldstone boson within the
top-quark triangle of the double-triangle seed diagram gives rise to a
${\cal O}(\alpha_s^2x_t)$ correction, which we must include in our analysis.
In the framework of the $n_f=5$ effective theory, where the top quark only
appears in the coefficient functions, this class of contributions is generated 
by the interference diagram of the operators $\left[O_1^\prime\right]$ and
$\left[O_{2q}^\prime\right]$ depicted in Fig.~\ref{figgammamix}.
The absorptive part of this diagram also includes a contribution from the $gg$
cut, which is well known and must be subtracted in order to obtain the desired
${\cal O}(\alpha_s^2x_t)$ correction to $\Gamma\left(H\to q\bar q\right)$.
We so obtain
\begin{equation}
\Xi_q^{\rm QCD}=\frac{\alpha_s^{(5)}(\mu)}{\pi}C_F
\left(-19+6\zeta(2)-\ln^2\frac{m_q^2}{M_H^2}-6\ln\frac{\mu^2}{M_H^2}\right).
\end{equation}
While $\Xi_q^{\rm QCD}$ is obviously not RG invariant, the physical observable
$\Gamma\left(H\to q\bar q\right)$ is because, in Eq.~(\ref{gammahqq}), the
$\mu$ dependence of $\Xi_q^{\rm QCD}$ is compensated by that of
${\cal C}_{2q}$, as will become apparent in the next section.

As a by-product, we may derive from the $n_f=5$ effective
Lagrangian~(\ref{eqlefffin}) a formula for the $H\to gg$ decay width
which includes the ${\cal O}(x_t)$ correction. 
Our result is
\begin{equation}
\label{gammahgg}
\Gamma(H\to gg)=A_{gg}\left(1+\bar\delta_{\rm u}\right)^2({\cal C}_1)^2,
\end{equation}
where
\begin{equation}
A_{gg}=\frac{N_cC_FG_FM_H^3}{\pi\sqrt{2}}.
\end{equation}
Notice that, to the order of our calculation, ${\cal C}_1$ is independent of 
the quark flavour.

At this point, we should mention that the QCD correction \cite{ina} to
$\Gamma(H\to gg)$ also includes contributions due to $q\bar qg$ final states.
These may also be interpreted as ${\cal O}(\alpha_s^3M_H^2/m_q^2)$ corrections
to the respective $H\to q\bar q$ decay widths \cite{pre,zer}.
Specifically, these corrections would appear in form of a term proportional to
$\left(\alpha_s^{(5)}/\pi\right)(M_H^2/m_q^2)\left({\cal C}_1\right)^2$
within the square brackets of Eq.~(\ref{gammahqq}).
In the following, this term will not be taken into account, since it is
formally one order of $\alpha_s$ beyond our considerations.

In the next section, we shall derive ${\cal C}_1$ as well as ${\cal C}_{2q}$
for $q\ne b$.
This will enable us to calculate $\Gamma\left(H\to q\bar q\right)$ through
${\cal O}(\alpha_s^2x_t)$ from Eq.~(\ref{gammahqq}) and
$\Gamma(H\to gg)$ through ${\cal O}(x_t)$ from Eq.~(\ref{gammahgg}).
The case $q=b$ is more complicated and will be treated in Sect.~\ref{sechbb}.


\section{$H\to q\bar q$ decay to ${\cal O}(\alpha_s^2x_t)$}
\label{sechqq}

\begin{figure}[ht]
   \leavevmode
 \begin{center}
 \begin{tabular}{ccc}
   \epsfxsize=4.0cm
   \epsffile[189 361 423 565]{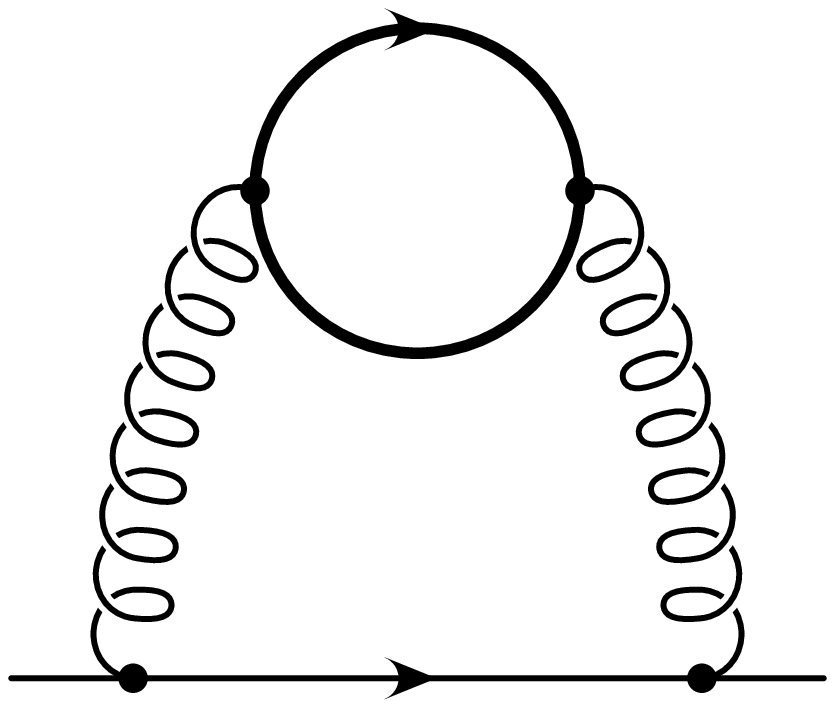}
&
   \epsfxsize=4.0cm
   \epsffile[189 361 423 565]{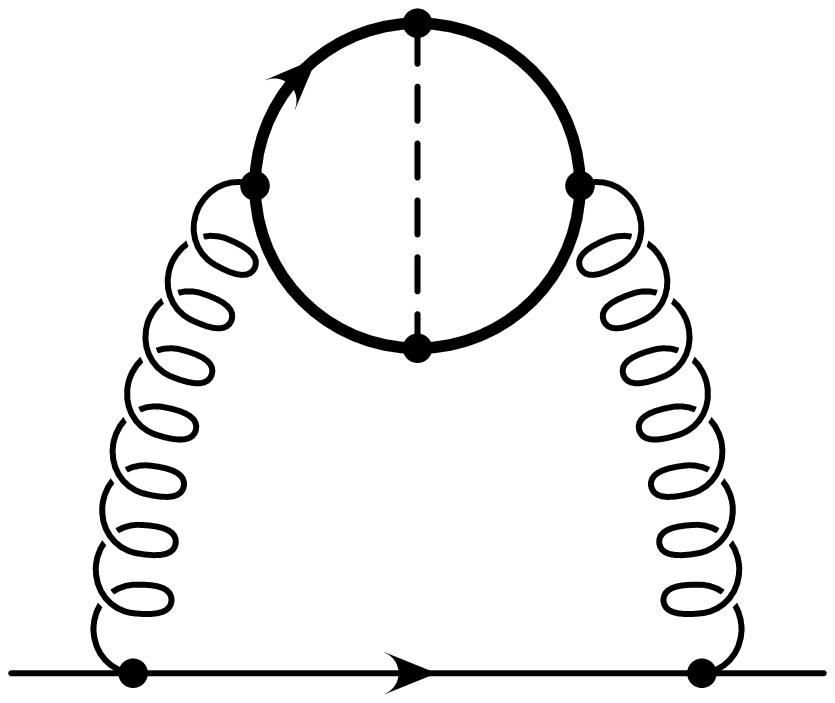}
&
   \epsfxsize=4.0cm
   \epsffile[189 361 423 565]{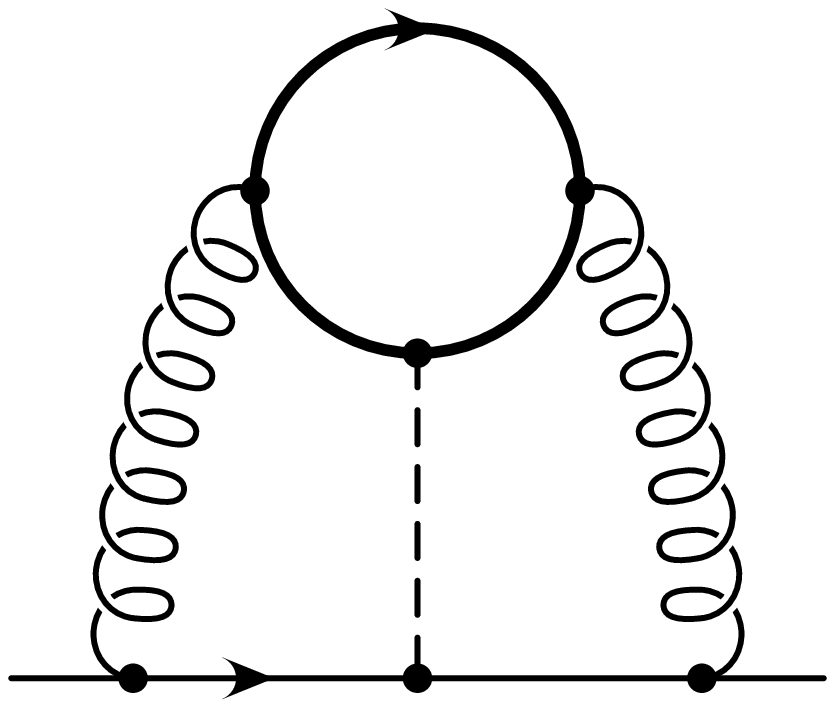}
\\
(a)&(b)&(c)
 \end{tabular}
  \caption{\label{fighqqdiag} Typical diagrams generating universal
${\cal O}(\alpha_s^2x_t)$ corrections to $\Gamma\left(H\to q\bar q\right)$.
The dashed line in diagram (b) represents a $H$, $\chi$, or $\phi^\pm$ boson,
while in diagram (c) only the $H$ and $\chi$ bosons appear.
In total, there are 12 diagrams of type (b) and 12 diagrams of type (c).}
 \end{center}
\end{figure}

In this section, we calculate the ${\cal O}(\alpha_s^2x_t)$ corrections
to the $H\to q\bar q$ decay widths, where $q\ne b$.
The relevant types of diagrams are depicted in Figs.~\ref{fighqqdiag}(a)--(c).
The diagrams of types (b) and (c) emerge from the pure QCD diagram in
Fig.~\ref{fighqqdiag}(a) by allowing for the exchange of one virtual Higgs or
Goldstone boson.
Although the pure QCD diagram only contributes in ${\cal O}(\alpha_s^2)$, it
is needed for the renormalization.
There, the top-quark mass appears logarithmically and has to be replaced
according to \cite{hem,boc}
\begin{equation}
m_t^0 = m_t\left(1+\frac{3}{2\varepsilon}x_t\right).
\end{equation}
The counterterm thus obtained cancels the ultraviolet subdivergences of the
type-(b) diagrams;
the remaining overall divergences are removed if the LET is applied.
We checked that this is true separately for the contributions due to the
$H$, $\chi$, and $\phi^\pm$ bosons.

The diagrams of type (c), where the scalar-boson lines are attached to both
the top-quark loop and the light-quark line, are also separately finite upon
application of the LET.
Due to the appearance of one power of the light-quark Yukawa coupling,
in our approximation, they only contribute to the scalar part of light-quark
self-energy, $\Sigma_S(0)$.
A potential problem arises in the diagrams of type (c) if the scalar particle
is the neutral Goldstone boson $\chi$, since then only one $\gamma_5$ matrix
appears in each quark line.
However, the arbitrariness of the $\gamma_5$ definition only affects the
finite terms of the original diagrams, which are removed by taking the
derivative according to the LET.

In the $\overline{\mbox{MS}}$ scheme, the coefficient functions are found to
be
\begin{eqnarray}
C_1 &\n=\n& \frac{\alpha_s^{(6)}(\mu)}{\pi}
             T\left(-\frac{1}{6}+\frac{1}{2}x_t\right),
\label{eqc1}
\\
C_{2q} &\n=\n& 1 + \left(\frac{\alpha_s^{(6)}(\mu)}{\pi}\right)^2
        C_FT\left\{\frac{5}{12}-\frac{1}{2}\ln\frac{\mu^2}{m_t^2}
              +x_t\left[\frac{7}{2}-18\zeta(3)+I_q\left(6+36\zeta(3)\right)
        \right]\right\},\quad
\label{eqc2q}
\end{eqnarray}
where $I_q$ is the third component of weak isospin, {\it i.e.},
$I_q=1/2$ for up-type quark flavours and $I_q=-1/2$ for down-type quark
flavours.
The dependence on $I_q$ stems from the $q\bar q\chi$ coupling, which appears
linearly in the $\chi$-exchange diagrams of type (c).
There are several checks for Eqs.~(\ref{eqc1}) and (\ref{eqc2q}).
If we ignore the RG improvement of Eq.~(\ref{eqctil}) and insert the
${\cal O}(\alpha_s)$ term of $C_1$ and the ${\cal O}(\alpha_s^2)$ term of
$C_{2q}$ into Eq.~(\ref{gammahqq}), then we recover the ${\cal O}(\alpha_s^2)$
double-triangle contribution to $\Gamma\left(H\to q\bar q\right)$ found in
Ref.~\cite{lar}.
The ${\cal O}(\alpha_sx_t)$ term of $C_1$ may be deduced from
Ref.~\cite{DjoGam94}.
The ${\cal O}(\alpha_s^2x_t)$ term of $C_2$ is new.
In compliance with RG invariance, it does not contain a logarithm of the type
$\ln(\mu^2/m_t^2)$.

Since the measured top-quark mass corresponds to the pole mass $M_t$, it is
convenient to directly express the perturbative expansions in terms of $M_t$.
In Eqs.~(\ref{gammahqq}) and (\ref{gammahgg}), the radiative corrections are
arranged in such a way that, in the limit $M_H\ll2M_t$, all dependence on the
top-quark mass is concentrated in $\bar\delta_{\rm u}$, ${\cal C}_1$, and
${\cal C}_{2q}$.
Consequently, we just need to substitute Eq.~(\ref{eqmbarmos}) into
Eqs.~(\ref{deltau}), (\ref{eqc1}), and (\ref{eqc2q}).
In the case of $C_{2q}$, this leads to
\begin{eqnarray}
C_{2q}^{\rm OS}&\n=\n&1+\left(\frac{\alpha_s^{(6)}(\mu)}{\pi}\right)^2
        C_FT\left\{\frac{5}{12}-\frac{1}{2}\ln\frac{\mu^2}{M_t^2}
        +X_t\left[\frac{15}{2}-18\zeta(3)+I_q(6+36\zeta(3))
\right.\right.\nonumber\\
&\n\n&{}+\left.\left.
\frac{3}{2}\ln\frac{\mu^2}{M_t^2}
\right]\right\},
\end{eqnarray}
where $X_t$ is defined below Eq.~(\ref{eqmbarmos}), while $C_1^{\rm OS}$
emerges from $C_1$ by merely replacing $x_t$ with $X_t$.
As for $\bar\delta_{\rm u}$, its counterpart in the on-shell scheme,
$\delta_{\rm u}$, may be found in Ref.~\cite{ks}.

We now turn to the $H\to gg$ decay width.
From Eq.~(\ref{gammahgg}) we obtain
\begin{equation}
\Gamma(H\to gg) = \frac{A_{gg}}{144}
              \left(\frac{\alpha_s^{(6)}(\mu)}{\pi}\right)^2
              (1+X_t),
\end{equation}
which is in agreement with Ref.~\cite{DjoGam94}.


\section{$H\to b\bar b$ decay to ${\cal O}(\alpha_s^2x_t)$}
\label{sechbb}

\begin{figure}[ht]
   \leavevmode
 \begin{center}
 \begin{tabular}{cc}
   \epsfxsize=4.0cm
   \epsffile[165 267 447 419]{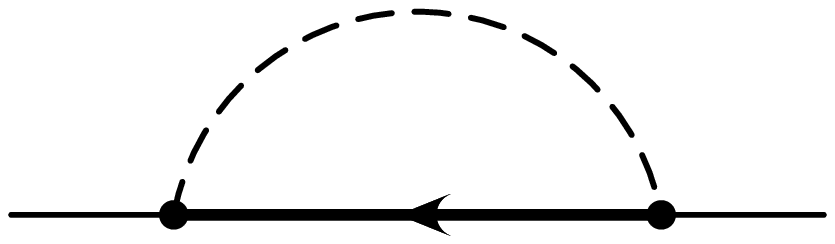}
&
   \epsfxsize=4.0cm
   \epsffile[165 267 447 419]{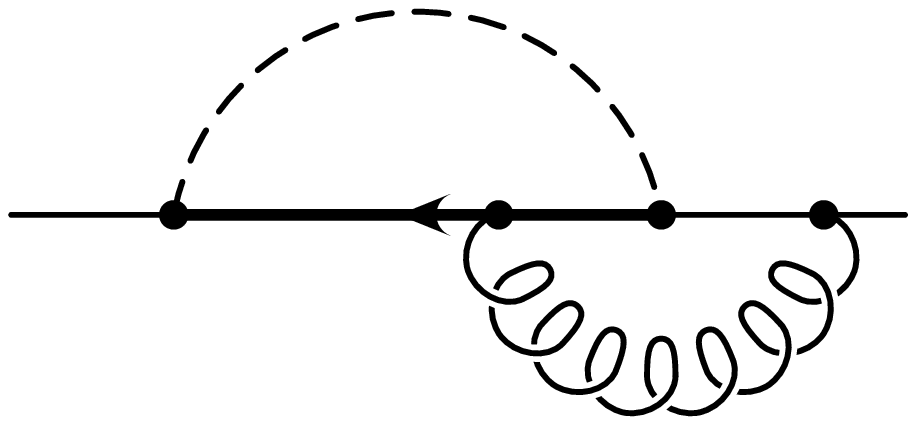}
 \end{tabular}
 \begin{tabular}{ccc}
   \epsfxsize=4.0cm
   \epsffile[165 267 447 419]{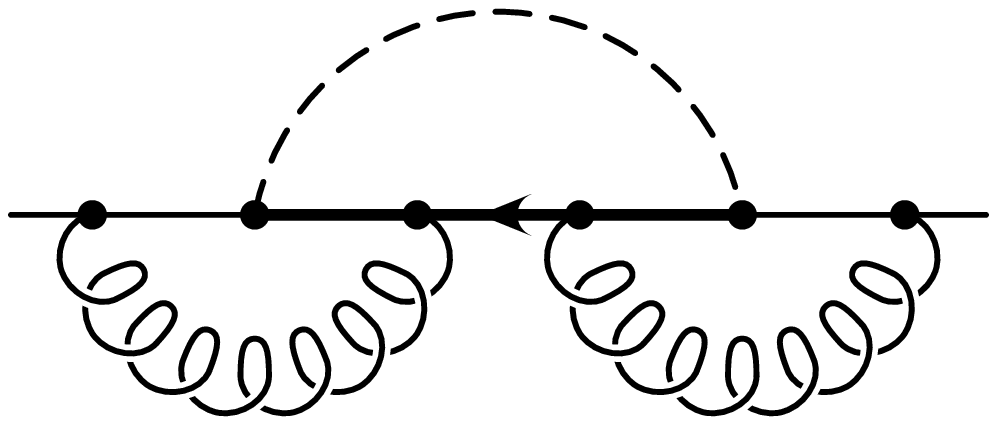}
&
   \epsfxsize=4.0cm
   \epsffile[165 267 447 419]{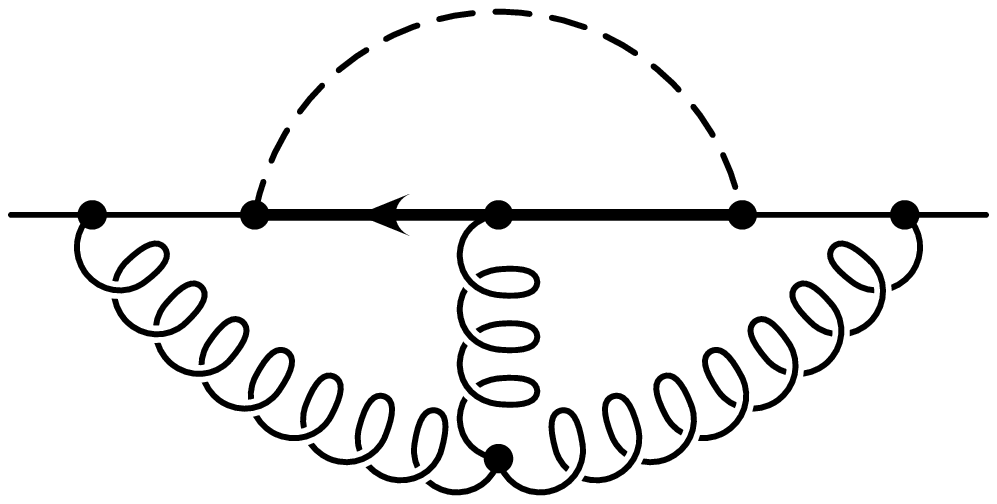}
&
   \epsfxsize=4.0cm
   \epsffile[165 267 447 419]{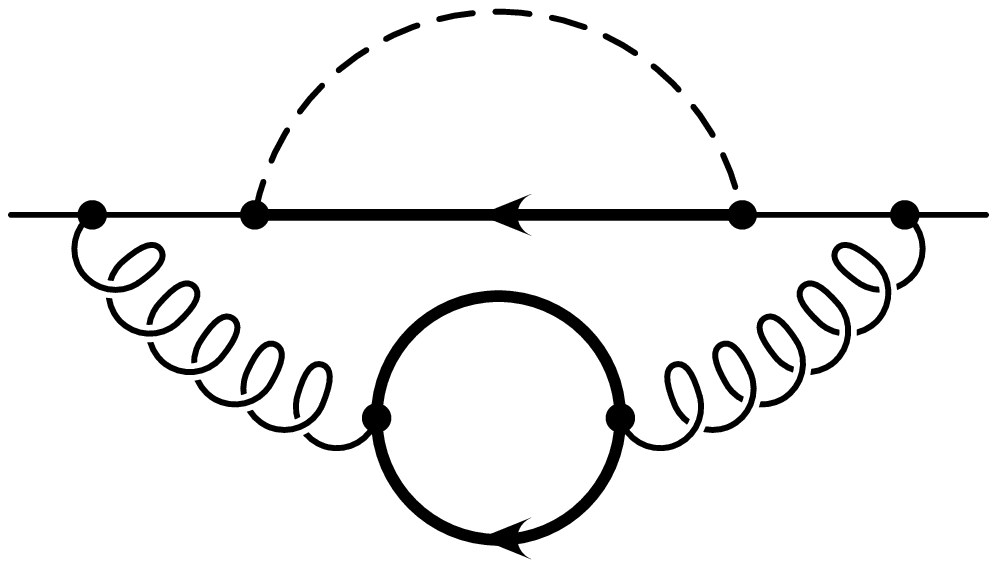}
 \end{tabular}
  \caption{\label{diahbbnu} Typical diagrams generating non-universal
${\cal O}(\alpha_s^nx_t)$ ($n=0,1,2$) corrections to
$\Gamma\left(H\to b\bar b\right)$.
The dashed lines represent the charged Goldstone boson $\phi^\pm$.
In total, there are 4 (54) irreducible two-loop (three-loop) diagrams.}
 \end{center}
\end{figure}

In this section, we upgrade the ${\cal O}(\alpha_s^2x_t)$ calculation of
$\Gamma\left(H\to q\bar q\right)$ to include the case $q=b$.
In addition to the diagrams in Fig.~\ref{fighqqdiag}, we must now consider 
diagrams where the primary bottom quark branches into a top quark and a 
charged Goldstone boson $\phi^\pm$.
Typical examples are depicted in Fig.~\ref{diahbbnu}.
At the three-loop level, there is a total of 54 such diagrams.
These diagrams only affect $C_{2q}$, while $C_1$ remains unchanged. 
Because the top-quark-induced corrections already start at ${\cal O}(x_t)$,
the renormalization constants $\zeta_{2q}^0$ and $\zeta_{m,q}^0$ of
Eqs.~(\ref{eqzeta2}) and (\ref{eqzetam}), respectively, do contribute to
Eqs.~(\ref{eqc2c3}) and (\ref{eqc3}), from which we extract $C_{2b}^0$.
Thereby, it is important to observe that the interference of ${\cal O}(x_t)$
and ${\cal O}(\alpha_s^2)$ terms contribute in ${\cal O}(\alpha_s^2x_t)$.

In the $\overline{\mbox{MS}}$ scheme, $C_{2b}$ is found to be
\begin{eqnarray}
\label{eqc2bms}
C_{2b} &\n=\n& C_{2d} + x_t\left\{
       -3
       +\frac{\alpha_s^{(n_f)}(\mu)}{\pi} C_F \left(-\frac{21}{4} 
                                    -\frac{9}{2}\ln\frac{\mu^2}{m_t^2} \right)
       +\left(\frac{\alpha_s^{(n_f)}(\mu)}{\pi}\right)^2
        \left[
         C_F^2 \left( 
                  - \frac{441}{64} 
\right.\right.\right.
\nonumber\\
&\n\n&{}+\left.
                   9\zeta(2) 
                  + \frac{99}{4}\zeta(3) 
                  - \frac{27}{16}\ln\frac{\mu^2}{m_t^2} 
                  - \frac{27}{8}\ln^2\frac{\mu^2}{m_t^2}
               \right)
       + C_F C_A \left(
                       -\frac{3629}{192} 
                       + \frac{11}{2}\zeta(2) 
                       - \frac{45}{8}\zeta(3)
\right.
\nonumber\\
&\n\n&{}-
\left.\left.\left.
                         \frac{87}{8}\ln\frac{\mu^2}{m_t^2} 
                       - \frac{33}{16}\ln^2\frac{\mu^2}{m_t^2}
                 \right)
       + C_F T n_f \left(    \frac{241}{48} 
                         - 2\zeta(2) 
                         + 3\ln\frac{\mu^2}{m_t^2} 
                         + \frac{3}{4}\ln^2\frac{\mu^2}{m_t^2}
                 \right)
\right]\right\}
\nonumber\\
&\n=\n& 
C_{2d} + x_t\left\{
-3 
+ \frac{\alpha_s^{(6)}(\mu)}{\pi}\left(-7 - 6\ln\frac{\mu^2}{m_t^2}\right) 
+ \left(\frac{\alpha_s^{(6)}(\mu)}{\pi}\right)^2\right.
\\
&\n\n&{}\times\left.
  \left[
     -\frac{3253}{48} 
     + 30\zeta(2) 
     + \frac{43}{2}\zeta(3)
     - \frac{69}{2}\ln\frac{\mu^2}{m_t^2} 
     - \frac{45}{4}\ln^2\frac{\mu^2}{m_t^2}
\right]
\right\},
\nonumber
\end{eqnarray}
where $n_f=6$.
Replacing $m_t$ by $M_t$ via Eq.~(\ref{eqmbarmos}) leads to
\begin{eqnarray}
\label{eqc2bos}
C_{2b}^{\rm OS} &\n=\n& C_{2d}^{\rm OS} + X_t
\left\{
  -3 
  +\frac{\alpha_s^{(n_f)}(\mu)}{\pi} C_F \frac{3}{4}
  +\left(\frac{\alpha_s^{(n_f)}(\mu)}{\pi}\right)^2
      \left[C_F T \left(
              - \frac{9}{2} 
              + 9\zeta(2)
          \right)\right.\right.
\nonumber
\\
&\n\n&{}
     +C_F^2 \left[
              -\frac{279}{32} 
              +9\zeta(2)\left(\frac{9}{4}-2\ln2\right) 
              +\frac{117}{4}\zeta(3)
            \right]
     +C_A C_F \left[
              -\frac{37}{24} 
              +\zeta(2)\left(\frac{5}{2}+9\ln2\right)
\right.
\nonumber
\\
&\n\n&{}-
\left.\left.\left.
               \frac{63}{8}\zeta(3)
              +\frac{11}{16}\ln\frac{\mu^2}{M_t^2} 
              \right) 
     +C_F T n_f \left(
                \frac{7}{12} 
              - 5\zeta(2)
              - \frac{1}{4}\ln\frac{\mu^2}{M_t^2}
              \right) 
   \right]
\right\}
\nonumber\\
&\n=\n& C_{2d}^{\rm OS} + X_t
\left\{
  -3 
  +\frac{\alpha_s^{(6)}(\mu)}{\pi} 
  +\left(\frac{\alpha_s^{(6)}(\mu)}{\pi}\right)^2
\right.
\nonumber\\
&\n\n&{}\times\left.
   \left[
        -\frac{67}{3} 
        +4\zeta(2)(8+\ln2) 
        +\frac{41}{2}\zeta(3)
        +\frac{7}{4}\ln\frac{\mu^2}{M_t^2}
\right]
\right\}.
\end{eqnarray}
To simplify Eqs.~(\ref{eqc2bms}) and (\ref{eqc2bos}), we set $N_c=3$, insert
the numerical values of the mathematical constants, and eliminate the RG
logarithms by choosing $\mu=\mu_t$ and $\mu=M_t$, respectively.
We so obtain
\begin{eqnarray}
\label{c2bnum}
C_{2b}&\n=\n&1
+\frac{5}{18}\left(\frac{\alpha_s^{(6)}(\mu_t)}{\pi}\right)^2
+x_t(\mu_t)\left[-3-7\frac{\alpha_s^{(6)}(\mu_t)}{\pi} 
+\left(\frac{\alpha_s^{(6)}(\mu_t)}{\pi}\right)^2
\right.\nonumber\\
&\n\n&{}\times
\left.\vphantom{\left(\frac{\alpha_s^{(6)}(\mu_t)}{\pi}\right)^2}
(-28.018+1.154\,n_f)\right],
\\
\label{c2bosnum}
C_{2b}^{\rm OS}&\n=\n&1
+\frac{5}{18}\left(\frac{\alpha_s^{(6)}(M_t)}{\pi}\right)^2
+X_t\left[-3+\frac{\alpha_s^{(6)}(M_t)}{\pi}
+\left(\frac{\alpha_s^{(6)}(M_t)}{\pi}\right)^2
\right.\nonumber\\
&\n\n&{}\times
\left.\vphantom{\left(\frac{\alpha_s^{(6)}(M_t)}{\pi}\right)^2}
(64.223-5.094\,n_f)\right],
\end{eqnarray}
where we have displayed the coefficient of $n_f=6$, for reasons which will
become clear in a moment.

Equation~(\ref{c2bosnum}) extends the non-universal correction
$\delta_{\rm nu}$ defined by Eq.~(12) of Ref.~\cite{hbb} to
${\cal O}(\alpha_s^2X_t)$.
From Eq.~(\ref{c2bosnum}) we read off that the leading ${\cal O}(X_t)$ term
receives the QCD correction factor
$\left[1-0.333\,\alpha_s^{(6)}(M_t)/\pi
-11.219\left(\alpha_s^{(6)}(M_t)/\pi\right)^2\right]$.
We thus recover a pattern similar to the electroweak parameter $\Delta\rho$
\cite{avd} and the corrections $\delta_{\rm u}$, $\delta_{WWH}$, and
$\delta_{ZZH}$ \cite{ks} to the $l^+l^-H$, $W^+W^-H$, and $ZZH$ vertices,
respectively.
In fact, the corresponding QCD expansions in $\alpha_s^{(6)}(M_t)/\pi$ of these 
four observables all have negative coefficients which dramatically increase in
magnitude as one passes from two to three loops \cite{ks}.
On the other hand, if the top-quark mass is renormalized in the 
$\overline{\mbox{MS}}$ scheme at scale $\mu=\mu_t$, then the respective QCD
expansions in $\alpha_s^{(6)}(\mu_t)/\pi$ are found to have coefficients which
have variant signs and nicely group themselves around zero \cite{ks}.
In the case of $C_{2b}$, the QCD correction factor reads
$\left[1+2.333\,\alpha_s^{(6)}(\mu_t)/\pi
+7.032\left(\alpha_s^{(6)}(\mu_t)/\pi\right)^2\right]$.
We conclude that, also in the case of the $b\bar bH$ interaction,
the QCD expansion in the on-shell scheme exhibits a worse convergence 
behaviour than the one in the $\overline{\mbox{MS}}$ scheme.
However, the difference is less striking than in the previous four cases.

Finally, we would like to test Broadhurst's rule concerning the na\"\i ve
non-abelianization of QCD \cite{bro}.
Guided by the observation that the $n_f$-independent term of $\beta_0$ in
Eq.~(\ref{beta}) emerges from the coefficient of $n_f$ by multiplication with
$-33/2$, Broadhurst conjectured that this very relation between the
$n_f$-independent term and the coefficient of $n_f$ approximately holds for
any observable at next-to-leading order in QCD.
In Ref.~\cite{ks}, this rule was applied to $\Delta\rho$, $\delta_{\rm u}$,
$\delta_{WWH}$, and $\delta_{ZZH}$, and it was found that, in all four cases,
the signs and orders of magnitude of the $n_f$-independent terms are correctly
predicted.
Except for $\delta_{ZZH}$, these predictions come, in fact, very close to the
true values.
If we multiply the coefficients of $n_f$ in Eqs.~(\ref{c2bnum}) and
(\ref{c2bosnum}) with $-33/2$, we obtain $-19.041$ and $84.055$, which has to
be compared with the respective $n_f$-independent terms, $-28.018$ and
$64.223$.
Once again, the signs and orders of magnitude of the $n_f$-independent terms,
which are usually much harder to compute than the coefficients of $n_f$, come
out correctly.


\section{Discussion and summary}
\label{seccon}

In this paper, we calculated the three-loop ${\cal O}(\alpha_s^2X_t)$ 
corrections to the $H\to q\bar q$ decay widths of the SM Higgs boson with mass
$M_H\ll2M_t$, including the case $q=b$.
To this end, we constructed a $n_f=5$ effective Yukawa Lagrangian by
integrating out the top quark.
This Lagrangian is a linear combination of dimension-four operators acting in
QCD with $n_f=5$ quark flavours, while all $M_t$ dependence is contained in
the coefficient functions.
We renormalized this Lagrangian and, by exploiting the RG invariance of the
energy-momentum tensor, rearranged it in such a way that the renormalized
operators and coefficient functions are separately independent of the
renormalization scale, $\mu$, to the order considered.
The RG-improved formulation thus obtained provides a natural separation of the
$n_f=5$ QCD corrections at scale $\mu=M_H$ and the top-quark-induced $n_f=6$
corrections at scale $\mu=M_t$, in the sense that the final result does not
contain logarithms of the type $\ln(M_t^2/M_H^2)$ if the $n_f=5$ and $n_f=6$ 
corrections are expanded in $\alpha_s^{(5)}(M_H)$ and $\alpha_s^{(6)}(M_t)$, 
respectively.

In contrast to the two-loop ${\cal O}(\alpha_sX_t)$ case \cite{hbb}, where it
was sufficient to consider just one term in the Lagrangian, we needed to take
into account three types of operators and to allow for them to mix under
renormalization.
The mixing terms are related to the ${\cal O}(\alpha_s^2)$ double-triangle
contribution considered in Ref.~\cite{lar} and extend the latter to
${\cal O}(\alpha_s^2X_t)$.

Similarly to Ref.~\cite{hbb}, we could take advantage of a low-energy theorem 
to simplify the calculation of the coefficient functions.
This allowed us to relate a huge number of three-loop three-point diagrams to
a manageable number of three-loop two-point diagrams.
Specifically, we had to compute 24 irreducible three-loop two-point diagrams
for $q\ne b$ and, in addition, 54 ones for $q=b$.
Such a theorem is not available for the gauge interactions, which might
explain why three-loop ${\cal O}(\alpha_s^2X_t)$ corrections have not yet been
calculated for the $Z\to q\bar q$ decay widths, including the important case of
$Z\to b\bar b$, which has recently attracted much attention in connection with
the so-called $R_b$ anomaly.

To illustrate the effect of the ${\cal O}(\alpha_s^2X_t)$ corrections to the
$H\to q\bar q$ decay widths, we rewrite Eq.~(\ref{gammahqq}) as
\begin{equation}
\Gamma\left(H\to q\bar q\right)=\Gamma_{q\bar q}^{\rm Born}
\left[\left(1+\Delta_q^{\rm QED}\right)
\left(1+\left.\Delta_q^{\rm weak}\right|_{x_t=0}\right)
\left(1+\Delta_q^{\rm QCD}\right)\left(1+\Delta_q^t\right)
+\Xi_q^{\rm QCD}\Xi_q^t\right],
\end{equation}
where
\begin{eqnarray}
\Delta_q^t&\n=\n&\left(1+\bar\delta_{\rm u}\right)^2\left({\cal C}_{2q}
\right)^2-1,
\nonumber\\
\Xi_q^t&\n=\n&\left(1+\bar\delta_{\rm u}\right)^2{\cal C}_1{\cal C}_{2q}
\end{eqnarray}
contain the leading $M_t$ dependence and refer to the $n_f=6$ theory, while
$\Delta_q^{\rm QCD}$ and $\Xi_q^{\rm QCD}$ are confined to pure QCD with
$n_f=5$.
We now collect our final results for $\Delta_q^t$ and $\Xi_q^t$.
For simplicity, we undo the RG improvement of Eq.~(\ref{eqctil}) and employ
the expanded versions of ${\cal C}_1$ and ${\cal C}_{2q}$ given in
Eq.~(\ref{calc}) instead.
In the $\overline{\mbox{MS}}$ scheme with $\mu=\mu_t$, we then have
\begin{eqnarray}
\Delta_u^t&\n=\n&
\left(\frac{\alpha_s^{(6)}(\mu_t)}{\pi}\right)^2\left(\frac{5}{9}
+\frac{2}{3}\ln\frac{\mu_t^2}{M_H^2}\right)
+x_t(\mu_t)\left[7+6.087\frac{\alpha_s^{(6)}(\mu_t)}{\pi}
+\left(\frac{\alpha_s^{(6)}(\mu_t)}{\pi}\right)^2
\right.\nonumber\\
&\n\n&{}\times\left.
\left(-6.640+\frac{8}{3}\ln\frac{\mu_t^2}{M_H^2}\right)\right],
\nonumber\\
\Delta_d^t&\n=\n&
\left(\frac{\alpha_s^{(6)}(\mu_t)}{\pi}\right)^2\left(\frac{5}{9}
+\frac{2}{3}\ln\frac{\mu_t^2}{M_H^2}\right)
+x_t(\mu_t)\left[7+6.087\frac{\alpha_s^{(6)}(\mu_t)}{\pi}
+\left(\frac{\alpha_s^{(6)}(\mu_t)}{\pi}\right)^2
\right.\nonumber\\
&\n\n&{}\times\left.
\left(-72.339+\frac{8}{3}\ln\frac{\mu_t^2}{M_H^2}\right)\right],
\nonumber\\
\Delta_b^t&\n=\n&
\left(\frac{\alpha_s^{(6)}(\mu_t)}{\pi}\right)^2\left(\frac{5}{9}
+\frac{2}{3}\ln\frac{\mu_t^2}{M_H^2}\right)
+x_t(\mu_t)\left[1-7.913\frac{\alpha_s^{(6)}(\mu_t)}{\pi}
+\left(\frac{\alpha_s^{(6)}(\mu_t)}{\pi}\right)^2
\right.\nonumber\\
&\n\n&{}\times\left.
\left(-59.163+\frac{2}{3}\ln\frac{\mu_t^2}{M_H^2}\right)\right],
\nonumber\\
\Xi_q^t&\n=\n&\frac{\alpha_s^{(6)}(\mu_t)}{\pi}
\left(-\frac{1}{12}-\frac{1}{3}x_t(\mu_t)\right),
\nonumber\\
\Xi_b^t&\n=\n&\frac{\alpha_s^{(6)}(\mu_t)}{\pi}
\left(-\frac{1}{12}-\frac{1}{12}x_t(\mu_t)\right),
\end{eqnarray}
where $q=u,d,s,c$.
Clearly, $\Delta_c^t=\Delta_u^t$ and $\Delta_s^t=\Delta_d^t$.
Introducing the top-quark pole mass $M_t$ and choosing $\mu=M_t$, we find the
equivalent expressions
\begin{eqnarray}
\Delta_u^{t,{\rm OS}}&\n=\n&
\left(\frac{\alpha_s^{(6)}(M_t)}{\pi}\right)^2\left(\frac{5}{9}
+\frac{2}{3}\ln\frac{M_t^2}{M_H^2}\right)
+X_t\left[7-12.580\frac{\alpha_s^{(6)}(M_t)}{\pi}
+\left(\frac{\alpha_s^{(6)}(M_t)}{\pi}\right)^2
\right.\nonumber\\
&\n\n&{}\times\left.
\left(-95.517+\frac{8}{3}\ln\frac{M_t^2}{M_H^2}\right)\right],
\nonumber\\
\Delta_d^{t,{\rm OS}}&\n=\n&
\left(\frac{\alpha_s^{(6)}(M_t)}{\pi}\right)^2\left(\frac{5}{9}
+\frac{2}{3}\ln\frac{M_t^2}{M_H^2}\right)
+X_t\left[7-12.580\frac{\alpha_s^{(6)}(M_t)}{\pi}
+\left(\frac{\alpha_s^{(6)}(M_t)}{\pi}\right)^2
\right.\nonumber\\
&\n\n&{}\times\left.
\left(-161.216+\frac{8}{3}\ln\frac{M_t^2}{M_H^2}\right)\right],
\nonumber\\
\Delta_b^{t,{\rm OS}}&\n=\n&
\left(\frac{\alpha_s^{(6)}(M_t)}{\pi}\right)^2\left(\frac{5}{9}
+\frac{2}{3}\ln\frac{M_t^2}{M_H^2}\right)
+X_t\left[1-10.580\frac{\alpha_s^{(6)}(M_t)}{\pi}
+\left(\frac{\alpha_s^{(6)}(M_t)}{\pi}\right)^2
\right.\nonumber\\
&\n\n&{}\times\left.
\left(-43.868+\frac{2}{3}\ln\frac{M_t^2}{M_H^2}\right)\right].
\end{eqnarray}
The expressions for $\Xi_q^{t,{\rm OS}}$ and $\Xi_b^{t,{\rm OS}}$ emerge 
from those of $\Xi_q^t$ and $\Xi_b^t$ by merely replacing $\mu_t$ and
$x_t(\mu_t)$ with $M_t$ and $X_t$, respectively.

At this point, we wish to emphasize that our calculation is only valid in the
limit $M_H\ll2M_t$, where the LET is applicable.
We expect that this is a good approximation for a Higgs boson in the
intermediate-mass range;
but, strictly speaking, this is just a conjecture, which can only be proven
once the calculation of the full mass dependence is available.
However, we can check the validity of this assumption at one loop, where the
radiative corrections are fully known \cite{fle,bak}.
Assuming $M_t=175$~GeV and $M_H=100$~GeV, we find that the ${\cal O}(X_t)$
term subtracted from $\left.\Delta_q^{\rm weak}\right|_{x_t=0}$ amounts to
81\% of the top-quark-induced correction inherent in $\Delta_q^{\rm weak}$.
It is reasonable to expect that this feature carries over to higher orders.

Notice that the branching ratios of the Higgs boson are more relevant 
phenomenologically than its partial decay widths,
especially in the intermediate-mass range, where the latter cannot be measured
directly.
In the branching ratios, the universal parts of the radiative corrections
cancel and, in fact, the only relevant correction is the one specific to the
$H\to b\bar b$ decay.

As a by-product of our analysis, we have confirmed a previous calculation
\cite{DjoGam94} of the ${\cal O}(X_t)$ correction to the $H\to gg$ decay
width.

\newpage
\centerline{\bf ACKNOWLEDGMENTS}
\smallskip\noindent
We thank J.H. K\"uhn for suggesting our collaboration on this project.
The work of one of us (M.S.) was partially supported by the
{\it Graduiertenkolleg Elementarteilchenphysik} of the University of 
Karlsruhe.


\end{document}